\documentclass[twocolumn,showpacs,prl]{revtex4}
\usepackage{graphicx}
\usepackage{dcolumn}
\usepackage{bm}

\begin{document}

\title{A scheme for demonstration of fractional statistics of anyons \\
in an exactly solvable model}
\author{Y.-J. Han$^{1}$, R. Raussendorf$^{2}$, and L.-M. Duan$^{1}$}
\address{$^1$ FOCUS center and MCTP, Department of Physics, University of Michigan, Ann
Arbor, MI 48109\\
$^2$ Perimeter Institute, Waterloo, Canada N2L 2Y5}

\begin{abstract}
We propose a scheme to demonstrate fractional statistics of anyons
in an exactly solvable lattice model proposed by Kitaev that
involves four-body interactions. The required many-body ground
state, as well as the anyon excitations and their braiding
operations, can be conveniently realized through \textit{dynamic
}laser manipulation of cold atoms in an optical lattice. Due to the
perfect localization of anyons in this model, we show that a quantum
circuit with only six qubits is enough for demonstration of the
basic braiding statistics of anyons. This opens up the immediate
possibility of proof-of-principle experiments with trapped ions,
photons, or nuclear magnetic resonance systems.
\end{abstract}

\maketitle

\address{1 FOCUS center and MCTP, Department of Physics, University of Michigan, Ann
Arbor, MI 48109\\
2 Perimeter Institute, Waterloo, Canada, ON, N2L 2Y5}

Anyons, as exotic quasiparticles living in two dimensions with
fractional statistics \cite{0}, have attracted strong interest
over the past two decades. The excitations in fractional quantum
Hall systems have been predicted to be anyons \cite{1}, and
fractional charges have been confirmed with some experimental
evidence \cite{evidence2}. However, a direct observation of
fractional statistics associated with anyon braiding is hard in
this system and has recently attracted intriguing theoretical
proposals \cite{3}. Other systems with anyon excitations have also
been proposed. In particular, in the context of topological
quantum computation, Kitaev described two exactly solvable
theoretical models \cite{K1, K2} which support anyons. An
implementation scheme has been proposed for the second model
Hamiltonian \cite{Duan, Zoller}, which involves only two-body
interactions. A detection method of anyons associated with this
implementation scheme has been proposed recently \cite{Zhang}.
This implementation scheme with an optical lattice requires
achievement of very low temperature \cite{Duan}, which
unfortunately is still somewhat beyond the current experimental
capabilities.

In this paper, we propose a method to realize anyons and to
observe their braiding statistics associated with the first Kitaev
model that is based on four-body interactions. Four-body
interactions are notoriously difficult to generate experimentally
in a controllable fashion. Our key idea here is to generate
\textit{dynamically} the ground state and the excitations of this
model Hamiltonian instead of direct ground-state cooling. Note
that the anyon properties are associated with the underlying
many-body entangled states. If we generate exactly the same ground
and the excited states, we should be able to observe the same
property. We propose several systems for implementation of this
idea. First, we show that laser manipulation of ultra-cold atoms
in an optical lattice provides a natural system to realize a large
scale ground state of this model Hamiltonian. The braiding statics
of anyons can then be observed with a series of single-bit
operations. Then, we show that the systems of trapped ions
\cite{5}, photons \cite{Pan}, or nuclear magnetic resonance
\cite{NMR}, provide a ready platform to realize a small scale
system for proof-of-principle demonstration of the anyon braiding
statistics. The anyons are perfectly localized quasiparticles in
this model Hamiltonian, which means we do not need a large system
for implementing their braiding operations. We show that as little
as six qubits are enough for demonstration of the basic braiding
statistics; and with nine qubits, one can further demonstrate
robustness of the braiding operation to certain variations of the
braiding path.

The first Kitaev model is a spin Hamiltonian for a two-dimensional
square lattice \cite{K1} (more generally, we can extend the model
from a square lattice to any planar graph). One associates each
edge of the lattice with a qubit (a spin-$1/2$ particle). The
model Hamiltonian is given by \cite{K1}
\begin{equation}
H=-\sum_{v}A_{v}-\sum_{f}B_{f},  \label{1}
\end{equation}
where $A_{v}$ is defined for each vertex $v$ as $A_{v}=\prod_{j\in
star(v)}X_{j}$, and $B_{f}$ is defined for each face $f$ as $%
B_{f}=\prod_{j=boundary(f)}Z_{j}$ (See Fig. 1A for an
illustration). The operators $X$ and $Z$ denote the standard Paul
matrix $\sigma _{x}$ and $\sigma _{z}$, respectively, and $A_{v}$
and $B_{f}$ are called the stabilizer operators in the context of
quantum error correction. In a square lattice, each term of the
Hamiltonian $H$ represents four-body interaction of local qubits.
One can easily check that all the terms of this Hamiltonian
commute, so it is straightforward to get the ground state $\left|
\varphi \right\rangle $ of the Hamiltonian $H$, which is given by
$A_{v}\left| \varphi \right\rangle =\left| \varphi \right\rangle $
and $B_{f}\left| \varphi \right\rangle =\left| \varphi
\right\rangle $\ for all vertexes and faces. The state $\left|
\varphi \right\rangle $ is highly entangled \cite{Zanardi}.

A quasiparticle is generated on the vertex $v_{0}$ (or face $%
f_{0}$) if $A_{v_{0}}$ (or $B_{f_{0}}$), acting on the excited state $\left|
\varphi _{e}\right\rangle $, yields an eigenvalue $-1$ instead of $+1$ for
the ground state. The quasi-particles in these two cases are called
e-particles (vertex) or m-particles (face), respectively. One can check that
the e-particles and m-particles by themselves are bosons, but the mutual
statistics between the e and m particles become the one for $1/2$-anyons
\cite{K1}, as we get a phase flip $e^{i\pi }$ if we move the e(m)-particle
around the m(e)-particle along an arbitrary loop.

It is hard to directly generate the interactions represented by the
Hamiltonian $H$ and to cool the system to its ground state $\left|
\varphi \right\rangle $. However, we note that the braiding
statistics of anyons are directly associated with the entanglement
properties of the underlying ground and excited states, whereas the
Hamiltonian only plays an implicit role for nailing down the
corresponding state. As long as we can create the state $\left|
\varphi \right\rangle $ and generate the e and m-particles above
this state, we should be able to demonstrate the fractional braiding
statistics between these quasiparticles. Our task reduces to how to
create the state $\left| \varphi \right\rangle $ and how to
demonstrate the fractional statistics of quasiparticles excited from
the state $\left| \varphi \right\rangle $. In the following, we
propose two approaches, targeted at a large scale implementation and
a small scale proof-of-principle demonstration, respectively.

Physically, the simplest method to generate a large scale state
$\left| \varphi \right\rangle $ is to start from a two-dimensional
(2D) cluster state. The cluster and the graph states, introduced in
\cite{HR1,R2}, are
defined as the co-eigenstate of a set of commuting stabilizer operators $%
S_{i}$ (with $+1$ eigenvalues). The 2D cluster state is associated
with qubits on vertexes of a 2D square lattice, where a stabilizer
operator is introduced for each vertex as $S_{i}=X_{i}\bigotimes_{j
\in N(i)}Z_{j}$. Therein, $N(i)$ denotes the set of nearest
neighbors of the vertex $i$. A 2D cluster state can be conveniently
generated with ultra-cold atoms in a square optical lattice. Each
atom is effectively two-level, which defines a qubit (or a spin-1/2
particle). One starts with all the atoms in equal superpositions of
$\left| 0\right\rangle $ and $\left| 1\right\rangle $ states
(co-eigenstates of $X_{i}$). Then, with control of the optical
lattice potential, one turns on spin-dependent collisions
\cite{Jaksch} or tunneling \cite{Duan} for a fixed amount of time
(both give rise to an effective Ising interaction) to get the 2D
cluster state \cite{HR1}. Starting from a 2D cluster state,
single-bit measurements of half of the qubits in the bases $X$ and
$Z$ respectively will yield \ the desired state $\left| \varphi
\right\rangle $ \cite{R3}. The measurement pattern is illustrated in
Fig.~1B. One can check
these measurements transfer the set of the stabilizer operators from $%
\left\{ S_{i}\right\} $ to $\left\{\pm A_{v}\text{,}\pm
B_{f}\right\} $ for the remaining qubits, as it is required for
the ground state of the Hamiltonian (1). The sign factors $\pm1$
in the stabilizer generators depend on the measurement outcomes
from the measured qubits. All these sign factors can be readjusted
to +1 by subsequent local Pauli operations.

\begin{figure}[tbp]
\includegraphics[width=8.0cm]{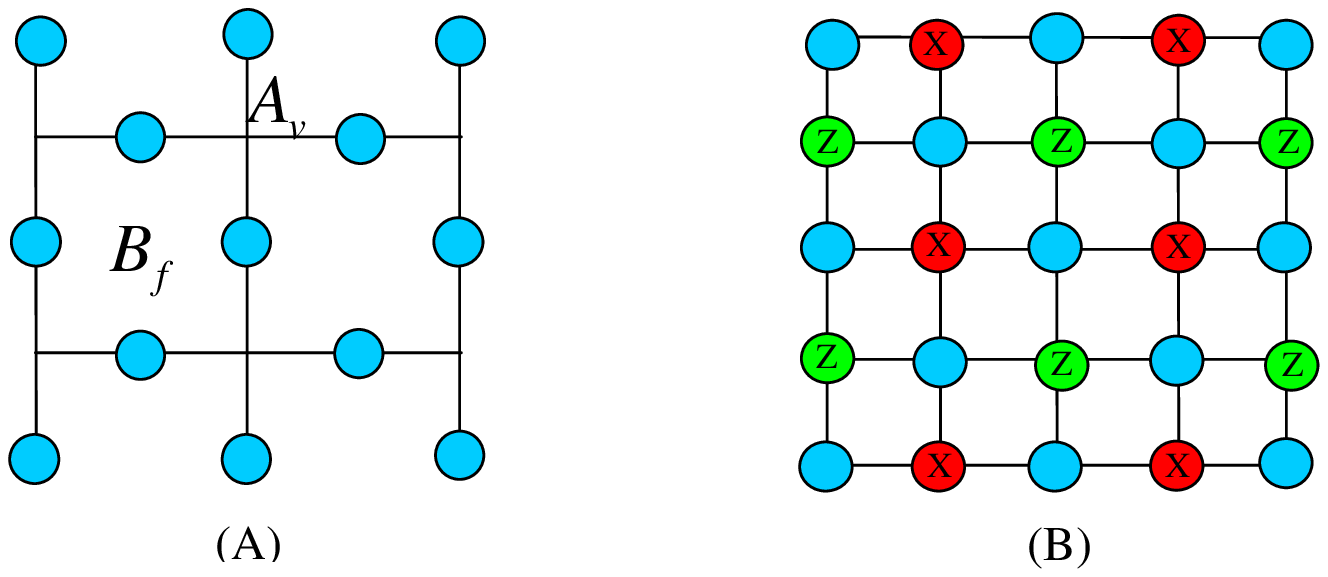} %
\includegraphics[width=8.0cm]{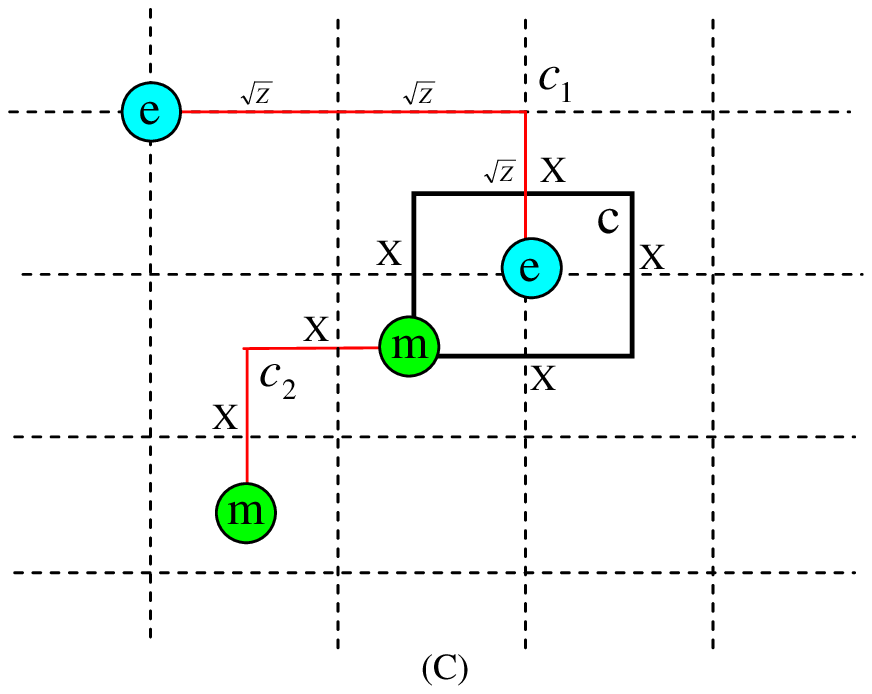}
\caption[Fig.I ]{(Color Online) (A)An illustration of the Kitaev
model Hamiltonian; (B) The measurement pattern to prepare the
many-body ground state $\left| \varphi \right\rangle $ from a
two-dimensional cluster state, where $X$ and $Z$ denote
respectively a single-bit measurement in the $X$ or $Z$ basis.
(see also \cite{R3}); (C) An illustration of braiding of anyons,
where a pair of e-particles (m-particles) are connected
respectively by the $c_1$ (or $c_2$) lines, and $c$ denotes the
braiding loop. The $X$ or $Z$ besides the edge qubits denote the
corresponding single-bit operations for creation and braiding of
the anyons. See the text for more detailed description.}
\end{figure}

With creation of the state $\left| \varphi \right\rangle $, one can
demonstrate the fractional statistical phase of the 1/2-anyons
through a Ramsey type interference experiment \cite{Zhang, BCZ}.
First, by applying a single-bit rotation $Z$ to one of the edge
qubits, one creates a pair of
e-particles on the neighboring vertexes, as represented by an excited state $%
\left| \varphi _{e}\right\rangle $ of the Hamiltonian (1). Then,
with half of the rotation $\sqrt{Z}$, we get the superposition
$\left( \left| \varphi \right\rangle +\left| \varphi
_{e}\right\rangle \right) /\sqrt{2}$. Similarly, with a single-bit
rotation $X$ on another edge qubit, we create a pair of m-particles
on the neighboring faces. Then, one can move one of the m-particles
around one of the e-particles along a loop through successive $X$
rotations acting on the qubits in the loop (see Fig. 1C). After
fusion of the m-particles (through another $X$ rotation), the
underlying state becomes $\left( \left| \varphi \right\rangle
-\left| \varphi _{e}\right\rangle \right) /\sqrt{2}$ due to the
fractional phase $\pi $ acquired from braiding of the e and
m-particles.
This phase flip can be unambiguously detected. For instance, with another $%
\sqrt{Z}$ operation, $\left( \left| \varphi \right\rangle -\left| \varphi
_{e}\right\rangle \right) /\sqrt{2}$ goes to $\left| \varphi \right\rangle $
(whereas without the statistics phase $\pi $, $\left( \left| \varphi
\right\rangle +\left| \varphi _{e}\right\rangle \right) /\sqrt{2}$\ would go
to $\left| \varphi _{e}\right\rangle $). The states $\left| \varphi
\right\rangle $ and $\left| \varphi _{e}\right\rangle $ can be distinguished
by measuring the relevant stabilizer operators $\left\{ A_{v}\right\} $
(measurement of the stabilizer operators only requires single-bit detection
together with classical correlation of the measurement outcomes).

In the above implementation scheme, except for the initial step of
preparation of the cluster state with spin-dependent atomic
collisions, all the following steps are based on single-bit
unitary operations or measurements. With cold atoms in an optical
lattice, one can indeed realize a large-scale system with millions
of qubits \cite{bloch}, and thus can well demonstrate the
robustness of the braiding operation: the statistics phase $\pi $
depends only on the topology of the loop and not on the detailed
path. The only remaining experimental challenge in this scheme is
the requirement of addressing of individual qubits for single-bit
operations. The experiments on ultracold atoms are making progress
towards the ability of single-bit addressing. On the other hand,
there are also other experimental systems which start from the
bottom up and have demonstrated the ability to fully control a
small number of qubits. Along with that line, in the following we
propose a different implementation scheme to provide
proof-of-principle demonstration of anyons and their braiding
statistics in small systems. This scheme is well within reach of
the current experimental technology.

It is interesting to ask what the smallest system size is for
demonstration of the braiding statistics of anyons. The system size
has to be significantly larger than the size of anyons for the
braiding operation. For any systems with anyons, the quasiparticles
are necessarily localized in space (otherwise it is impossible to
define the braiding). For the first Kitaev model, the quasiparticles
are particularly well localized: the e--particle is on a single
vertex and the m-particle is on a single face. Due to this perfect
localization, we do not need a large system for a basic braiding
operation. It turns out that the graph shown in Fig. 2A is the
smallest system for implementation of the anyon braiding operation,
which just requires manipulation of six qubits.

The graph in Fig. 2A corresponds to the ground state of the Hamiltonian $%
H_{6}=-A_{1}-A_{2}-B_{1}-B_{2}-B_{3}-B_{4}$ of six edge qubits, where $%
A_{1}=X_{1}X_{2}X_{3}$, $A_{2}=X_{3}X_{4}X_{5}X_{6}$, $B_{1}=Z_{1}Z_{3}Z_{4}$%
, $B_{2}=Z_{2}Z_{3}Z_{5}$, $B_{3}=Z_{4}Z_{6}$, $B_{4}=Z_{5}Z_{6}$.
One can check that its ground state $\left| \varphi \right\rangle
_{6}$ is equivalent under local single-bit operations to a graph
state shown in Fig.~2B (see Ref. \cite{R2} for definition of the
graph state). One can use five controlled phase flip (CPF) gates
to prepare the graph state of Fig.~2B and thus also $\left|
\varphi \right\rangle _{6}$. Figure 2C shows the detailed
preparation circuit for the state $\left| \varphi \right\rangle
_{6}$ which involves five CPF gates as well as a few Hardmard (H)
gates. Then, as also shown in Fig. 2C, a $\sqrt{Z_{3}}$ operation
on the qubit 3 generates the superposition state $\left( \left|
\varphi \right\rangle _{6}+\left| \varphi _{e}\right\rangle
_{6}\right) /\sqrt{2}$, where $\left| \varphi _{e}\right\rangle
_{6}$ has a pair of e-particles on two vertexes of the edge 3.
Another $X_{4}$ operation on the qubit 4 further generates a pair
of m-particles on two neighboring faces of the edge 4. With four
$X$ operations on the qubits 6-5-3-4, we achieve the braiding by
moving an m-particle around a e-particle along the loop shown in
Fig. 2A. Finally, we fuse (annihilate) all the quasiparticles (with another $%
\sqrt{Z_{3}}$ and $X_{4}$ operation), and the resultant state is detected
with single-bit measurements (in $X$ or $Z$\ basis) of the six stabilizer
operators.

\begin{figure}[tbp]
\includegraphics[width=8.0cm]{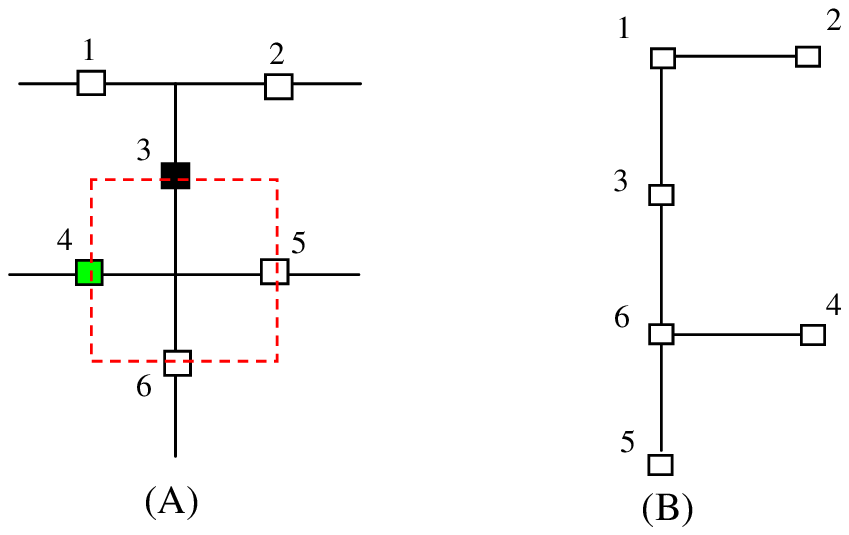} %
\includegraphics[width=8.5cm]{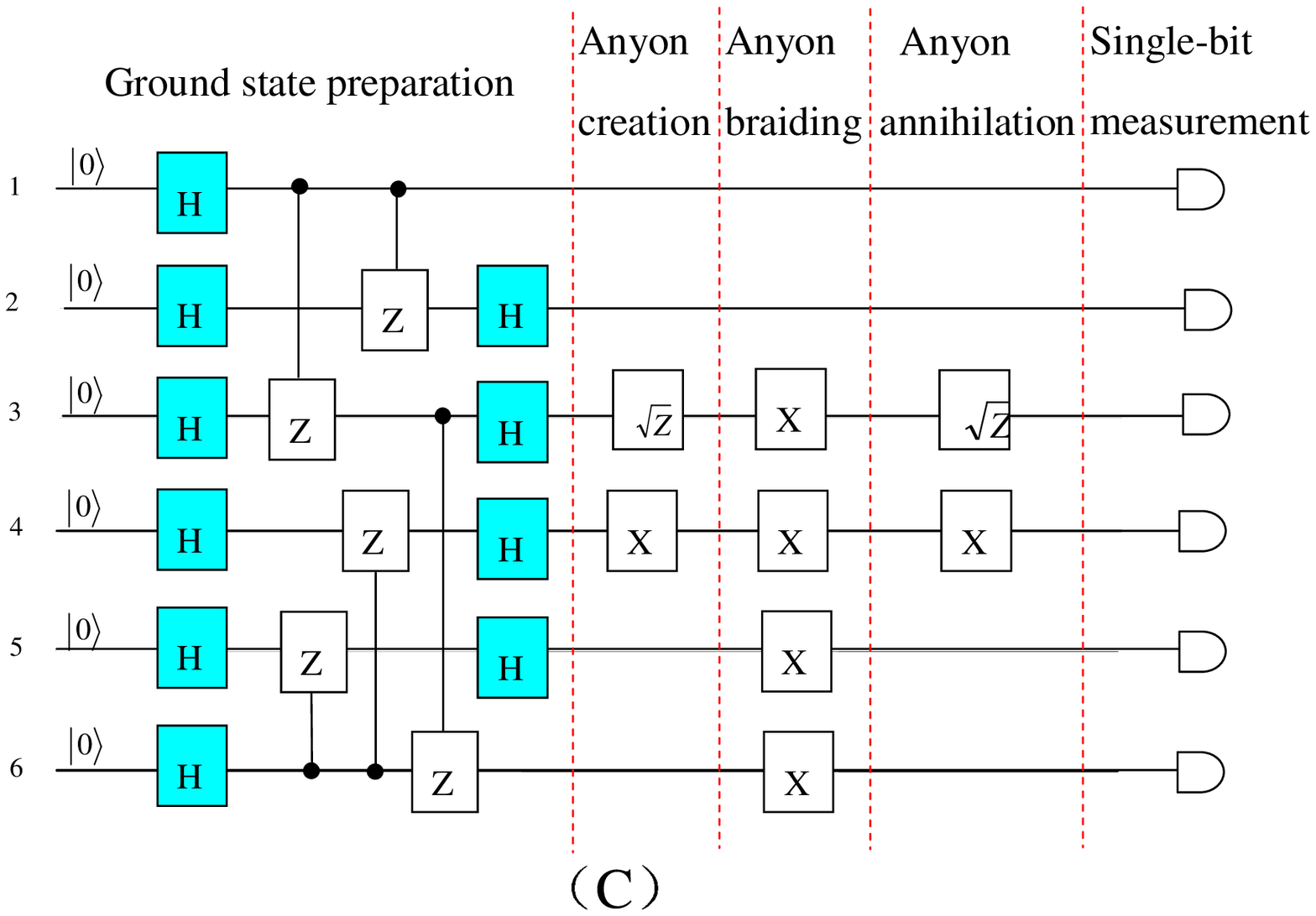}
\caption[Fig.III ]{(Color Online) (a) A planar graph
representation of the Hamiltonian with a minimum number of qubits
that support braiding of anyons. (B) The graph state that is
equivalent under local unitary operations to the ground state of
the Hamiltonian in Fig. 2A. (C) The detailed quantum circuit for
generating the ground state $\left| \varphi \right\rangle _{6}$,
for creation, braiding, and annihilation of the anyons, and for
the final state detection.}
\end{figure}

In the above six-qubit minimum implementation, the braiding loop
is unique and there is no way to show the robustness of the
topological braiding operation by moving the anyons along
different paths. To have demonstration of some robustness of the
anyon operation to different braiding paths, we give another
implementation in Fig. 3 which uses nine qubits. The ground state
of this nine-qubit Hamiltonian represented by Fig. 3A is locally
equivalent to graph state shown in Fig.~3B, and the detailed
implementation circuit is given in Fig.~3C. In this plane geometry
(see Fig.~3A), there are several different loops for the
quasiparticles. The anyonic operation only depends on the
topological character of the loop, i.e., whether the different
types of quasiparticle (e and m) ``braid'' with each other.  For
instance, both the loops 6-5-3-4 and 9-8-5-3-4-7 of the
m-particles (the dashed lines in Fig. 3A) give the same result of
the statistical phase, while the loop 9-8-6-7 (the solid line in
Fig 3A) has no effect since it does not enclose a e-particle.
Braiding along different loops can be implemented with single-bit
$X$ operations on different sets of individual qubits. See Fig.~3C
for an example circuit implementing braiding of a m-particle with
an e-particle. The m-particle is taken along the loop 9-8-5-3-4-7.
The enclosed e-particle resides on the vertex between edges
3,4,5,6.

\begin{figure}[tbp]
\includegraphics[width=8.0cm]{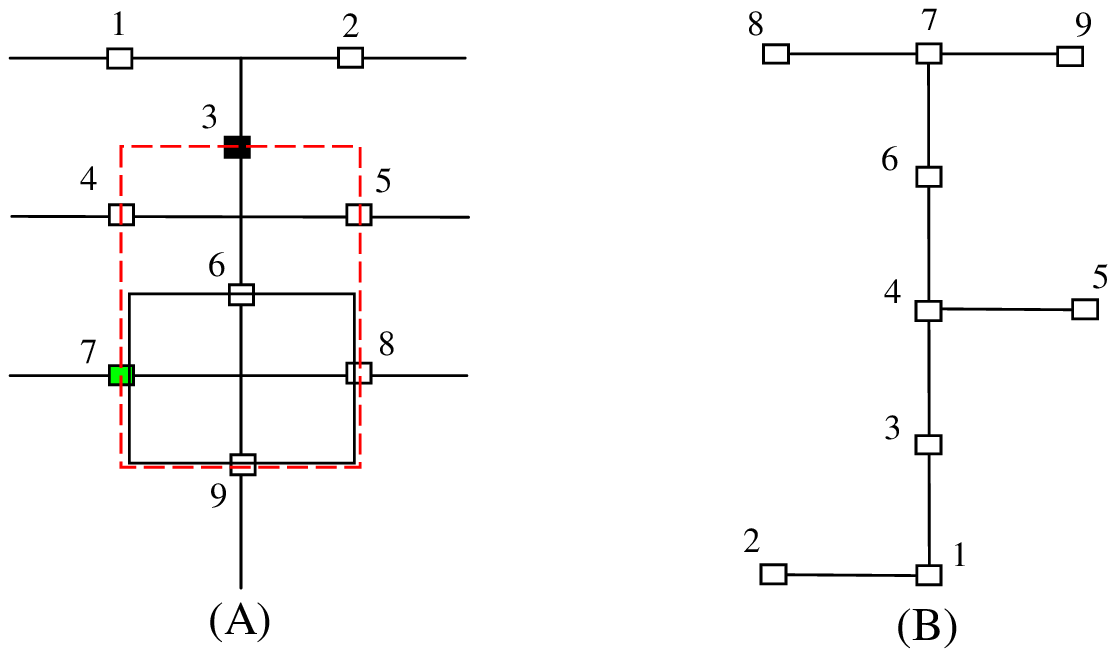} %
\includegraphics[width=8.5cm]{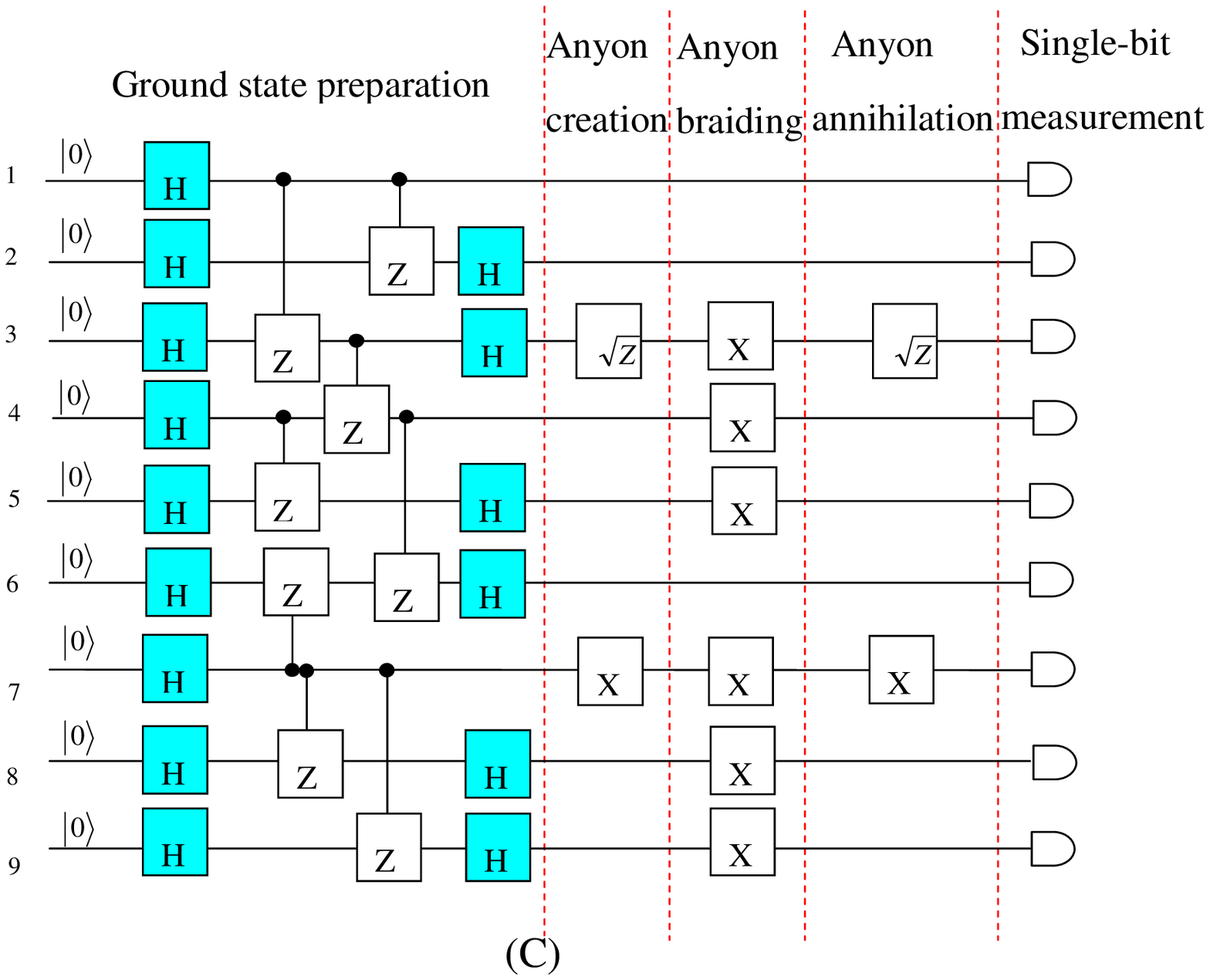}
\caption[Fig.IV ]{(Color Online) (A) A planar graph
  representation of the Hamiltonian with $9$ qubits that support
different types of loops for anyons. The dashed line loop denotes an
effective braiding while the solid line loop is not. (B) The graph
state locally equivalent to the ground state of the Hamiltonian in
Fi.g 3A. (C)The detailed quantum circuit for realization of the
ground state and the associated anyon braiding (the braiding
corresponds to the dashed line loop in Fig. 3A).}
\end{figure}

In the context of quantum computing, entangling gates of six to
eight qubits have been demonstrated in with trapped ions \cite{5},
six-photon entanglement in the linear optics system \cite{Pan},
and manipulation of a dozen of nuclear spins in the nuclear
magnetic resonance system \cite{NMR}. With such experimental
capabilities, the above six-bit and nine-bit implementation
schemes can be expected to be realized in the very near future.
This kind of proof-of-principle demonstration of anyons in small
and relatively simple systems will represent an important step
towards the long pursued goal to demonstrate fractional statistics
of quasiparticles in a macroscopic material. Such abilities,
properly extended to large systems, will also be critical for
future implementation of fault-tolerant topological quantum
computation.

In summary, we have proposed a method to demonstrate fractional
braiding statistics of anyons in an exactly solvable spin model. Two
types of implementation schemes are described, respectively targeted
at a large-scale realization with cold atoms in an optical lattice
and a small-scale realization with some available qubit systems,
such as trapped ions, photons, or nuclear spins in liquids of
molecules. The proposed schemes open the prospect of experimental
demonstration of anyons in the near future.

This work was supported by the NSF awards (0431476), the ARDA under
ARO contracts, the A. P. Sloan Fellowship and at the Perimeter
Institute of Theoretical Physics by the Government of Canada through
NSERC and by the Province of Ontario through MRI.

\end{document}